\documentclass{article}
\usepackage{arxiv}
\usepackage[utf8]{inputenc} 
\usepackage[T1]{fontenc}    
\usepackage{lineno,hyperref}
\usepackage{amssymb}
\usepackage{amsmath}

\usepackage{csquotes}
\usepackage{array}
\usepackage{tikz}
\def\checkmark{\tikz\fill[scale=0.25](0,.35) -- (.25,0) -- (1,.7) -- (.25,.15) -- cycle;}
\usepackage{url}            
\usepackage{booktabs}       
\usepackage{amsfonts}       
\usepackage{nicefrac}       
\usepackage{microtype}      
\usepackage{lipsum}
\graphicspath{ {images/} }
\title{\emph{InfoRest}: Restricting Privacy Leakage to Online Social Network App}

\author{
  Nemi Chandra Rathore
   \\
  Department of Computer Science \& Engineering\\
  Indian Institute of Technology Patna, Bihar\\
  \texttt{nemi.rathore@gmail.com} \\
   \And
Somanath Tripathy \\
  Department of Computer Science \& Engineering\\
  Indian Institute of Technology Patna, Bihar\\
  \texttt{som@iitp.ac.in} \\
}

\begin{document}
\maketitle

\begin{abstract}
In recent years, Online Social Networks (OSNs) have become immensely popular as social interaction services among worldwide Internet users. OSNs facilitate Third-party applications (TPAs) which provide many additional functionalities to users. While providing the extended services TPAs access the users data which would raise serious concerns to user privacy. This is due to the lack of user data protection mechanisms by OSN and none of the present OSN platforms offers satisfactory protection mechanisms to users private data. In this paper, we propose an access control framework called InfoRest to restrict user data to TPA in OSN, considering users privacy preferences and attribute generalization. Further, we propose a relation based access control (ReBAC) policy model and use predicate calculus to represent access conditions. The usability and correctness of the proposed policy model are demonstrated with the help of a logical model developed using answer set programming.     
\end{abstract}


\section{Introduction}\label{sec:intro}
Online Social Network (OSN) services have become popular in recent years with the increased penetration of Internet, and availability of low cost smart phones. As per the report from \textit{www.wearesocial.com} and \textit{Hootsuite}\footnote{www.hootsuite.com}, more than $50\%$ of world's population has access to Internet, and around $78\%$ of them are active social network users \cite{WeareSocial2017}. OSNs like Facebook\footnote{www.facebook.com}, Twitter\footnote{www.twitter.com}, Instagram\footnote{www.instagram.com} are some of the popular OSN services that have huge number of registered users (subscribers).
\par OSN services enable their subscribers to communicate with their social connections. They also allow their subscribers to form connections with other users on the basis of various common attributes like interests, geographic location, organization and so on. During interaction, subscribers  share and access variety of contents like text, photos, animations, and urls, that carry variety of personal information. Such a huge amount of personal information in OSN attract hackers and other malicious entities, that might cause variety of threats to the privacy of users \cite{Eecke2010}\cite{Carolyn2012}.  Unfortunately, existing solutions offered by OSNs for protecting user privacy are not satisfactory enough, and most of the times, users contents become available to unintended users resulting in privacy violation \cite{Joshi2011} \cite{Kayes2017}.
A number of methods for controlling the access to users private data from other users have been proposed  in \cite{Carminati2010}\cite{Hu2013}\cite{Cheng2012}\cite{Cheng2014}\cite{Rathore2017}.
\par OSNs like Facebook allow a wide range of \textit{Third Party Applications} to provide additional services and functions to make their services more attractive. Majority of these applications include online games and quizzes. These applications require access to some of the data of subscribers  to offer them customized services. These applications follow \textit{all or none approach}, that is, either user has to permit access to all the requested data to proceed for installation, or discontinue the installation of the app.
 \par All of these applications are developed and hosted by various untrusted third-party servers, which may cause different privacy risks to OSN users. For example, the investigation report published in \textit{Wall Street Journal} in October 2010\cite{Beaumont2010}, claimed that, the top ten Facebook apps such as \textit{FarmVille} and \textit{Mafia Wars} shared Facebook IDs, names and friend list of millions of Facebook users illegally with some of the advertising and tracking companies. Recently, following the famous \textit{Cambridge Analytica Scandal}, Facebook suspended 200 third-party applications that were found misusing users private data\cite{Levin2018}.  On the other hand, now a days some of the companies including OSN service providers, trade in currency of personal information \cite{Wong2018}, and allow various \textit{advertisers} and \textit{user tracking sites} to access users data that makes problem more severe\cite{Krishnamurthy:2009} \cite{Eckersley2009}\cite{Tomy2016}.  Therefore, there is an urgent need to develop a mechanism to regulate the access of user data by third-party entities.
   \par  The major contribution of this work is in two folds. First, we propose a new framework named \textit{InfoRest} to preserve  user privacy, by blocking flow of private data between internal and external modules of third party applications. Then, we propose a (Relation-based Access control) \textit{ReBAC policy model} to control access of user data by TPA components. One of the most attractive features of {\em InfoRest} is to facilitate the users to share \textit{generalized values} of their  private data to TPA for restricting their privacy leakages.
 \par This paper is organized into total 7 sections. We provide a brief review of related work in Section-2 and background of the work in Section-3. In Section-4, we present our proposed framework for OSN and an access policy model in Section-5. In Section-6, we discuss our framework in the light of available similar methods and finally, conclude our work in Section-7. 
 \section{Related Work}\label{sec:relwork}
 Research about privacy preservation on OSN platforms is in its early stage. A number of mechanisms to protect access of data to TPAs have been proposed in many works \cite{Singh:2009}\cite{Viswanath2012}\cite{Anthonysamy:2012}\cite{Egele2012}\cite{Giffin:2012}\cite{Shehab2012}\cite{Cheng2013}. In \cite{Singh:2009}, authors proposed a trusted TPA hosting platform named as xBook in order to preserve user privacy. It enforces flow control over the information flow at xBook platform as per the privacy policies.  It assumes that at the time of deployment, the TPA developer specifies about the private data each component consumes, and the external entities it communicates with. On the basis of this information, xBook creates a manifest that is used to enforce what data a component communicate with external entities and what user data is accessed by it. 
 \par Viswanath et. al. \cite{Viswanath2012} have proposed a sandbox model inspired by \textit{XBook} \cite{Singh:2009} which requires a TPA to be hosted on a trusted infrastructure containing two sandboxes. Both the application and user agent run in cloud, and the application accesses user data by querying the application's global database. The results of the query are served while assessing the information loss incurred and the privacy budget. The drawback of both the methods are that they require significant changes in the current architecture of existing OSNs. In \cite{Anthonysamy:2012}, authors proposed a collaborative model that uses an interceptor to capture all the data requests made by a TPA. The interceptor imposes a \textit{user configuration} on these requests. The user may share this configuration to other users as well which makes access control task easy for a novice user. But, the sharing of privacy preferences in the form of configuration also leaks users private information to their respective neighbors.
 \par In \cite{Egele2012}, authors have developed an extension to Facebook called \textit{PoX} to restrict the access of user data from TPAs. PoX has been developed as a third-party application to Facebook based on \textit{Client-Server} architecture. The client-side is implemented as a proxy between Facebook and a TPA. All the requests made by the TPA to Facebook are evaluated by PoX  looking at the subscribers access policies for TPA. Only the user access requests granted by the subscriber would be forwarded to Facebook.  PoX blocks all the requests to the data items that are not allowed by the policies of the subscriber. Giffin et. al. in \cite{Giffin:2012} proposed a new web framework named as Hails, that employs Mandatory Access Control (MAC) and a declarative policy language for specification of the policies. Shehab et. al. in \cite{Shehab2012} proposed a solution that models application as a Finite State Machine where execution of the application is conditioned on user data. It requires user to specify a minimum set of attributes and their minimum generalization levels to get specific services offered by the application.
 \par In \cite{Cheng2013}, authors have proposed a framework that exploits relationship between user and application for access control policy specification. This framework classifies TPAs into three categories, one that runs totally inside OSN, other that runs outside OSN and the third one whose some modules run at OSN site and remaining runs at outside servers. The idea behind this framework is that the private data should only be accessed at OSN site. Authors also proposed a relationship based policy specification language which exploits relationship between users and TPA. The drawback of this model is that, it assumes that the functional behavior of all internal components of a TPA is verified by OSN provider. It is difficult to ensure this, since it may lead to different copyright issues. Moreover, the policy model offered by this model is also not expressive enough.
 \section{Background}\label{sec:background}
  
 \par
\begin{figure}[!ht]
 	\begin{center}
 		\includegraphics[scale=0.3]{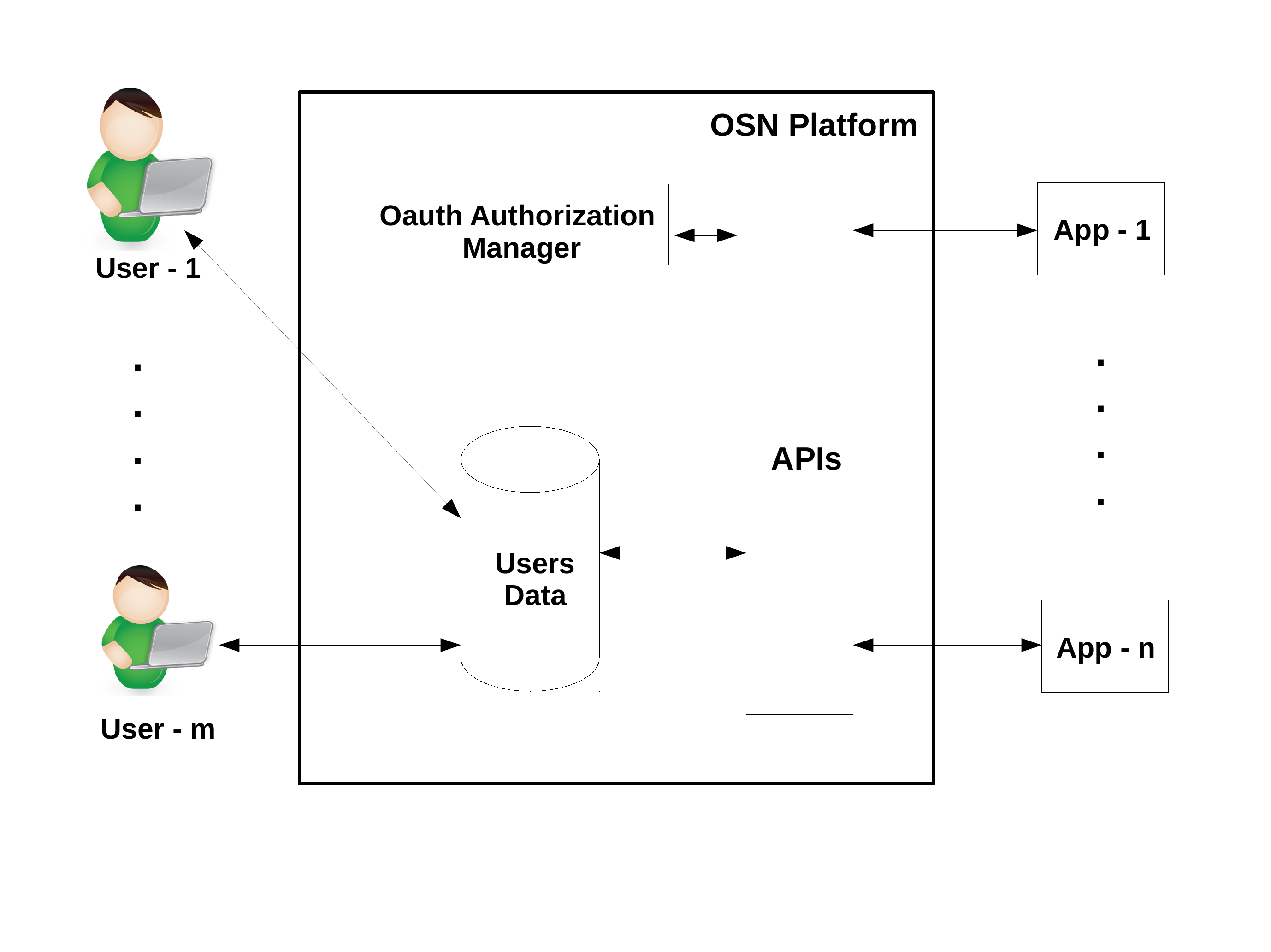}\label{fig:TraditionalTPA}
 		\vspace{-1.4cm}
 		\caption{Traditional OSN Architecture with TPA}\label{fig:trad_arch}
 	\end{center}
 \end{figure}

 Most of the popular OSNs allow TPA developers to implement variety of attractive services in OSN to increase their user base. To facilitates development of such applications an OSN offers a set of APIs. These APIs allows TPAs to communicate with the OSN users and accessing their data through.  The figure-\ref{fig:TraditionalTPA} shows a typical, OSN architecture for facilitating TPAs. 
 
  A third-party application on OSN requires permissions to access resources of its subscribers which might include private information of the subscribers. The TPA provider would access these private information and use maliciously by providing those to tracking companies or other similar entities. This creates a significant risk to the privacy of TPA subscribers which need to be addressed without affecting the core functionality of the third-party applications. In this section, we first pose the security requirements and the associated threat model. Subsequently, TPA installation, access to user data by TPA according to OAuth 2.0\cite{rfc6749}, are presented precisely. 
 
\subsection{Threat Model}\label{subsec:threatmodel}
At OSN platform, a malicious application $\mathcal{A}$, developed by an untrusted third-party developer $\mathcal{D}$ or be hosted at an untrusted third-party server $\mathcal{S}$. Traditionally, $\mathcal{A}$ requests access to a subset of private and public data from its subscribers  and offers them a set of customized services in return. 
$\mathcal{A}$ is comprised with a set of components  $\{C_1, C_2,...,C_n\}$. Some of these components might be residing on OSN site while others could be on the corresponding third-party server. Some Components of $\mathcal{A}$ target to leak private data of OSN users and access unauthorized data of its subscribers. We assume that the OSN server is trusted and does not leak any private information of its users. 
\subsection{Security Requirements }\label{subsec:secreq}
 \par {A TPA needs access to some personal information of OSN users, in order to customize its offered services for them. Once  the subscribers release their data to the TPA, they loose control on the data. Furthermore, they do not have any mechanism for determining the \textit{minimum set of attributes} needed by the TPA. Thus, following are the security requirements that any solution to the above said problem should conform:}
 \begin{enumerate}
 	\item \emph{Principle of least privileges:} A TPA must be given access to the minimum set of user data i.e. it should get access of only those user resources that are really necessary for the application to provide its services to the user.
 	\item \emph{Purposeful information flow:} User data should not be available to any component of the application which do not require it.
 	\item \emph{Zero Store:} An application should neither store nor communicate any private information of its subscribers outside the OSN. This is very important requirement because once data is sent to a third-party server, then both,  its owner and the OSN loose control over it.
 \end{enumerate}
  \subsubsection{Other Desirable Features}
  {Additionally, for a good solution for the posed problem, it is desirable to have the following features:}
  \begin{enumerate}
     \item {\textit{Transparency:} The mechanism should be transparent to both, the users and TPA. Further, it should not add significant delay to the requests made by a TPA, because large response time can make the TPA to loose its customers.}
      \item {\textit{Compatibility:} The mechanism should be able to fit easily with the existing OSN architecture. A solution requiring significant changes into the existing architecture would be difficult to adopt.}
      \item {\textit{User-TPA Access Control Policy Model:} Users should be able to regulate the data accessed by TPAs as per their privacy preferences. The policy language should be fine grained and expressive enough to express users privacy preferences. Furthermore, the policy model must include a \textit{conflict resolution mechanism} to deal with a \textit{policy conflict}. }
      \item {\textit{Components based TPA Model:}
      The TPA should be developed as a set of small modules or components. It makes the task of privacy preservation easy. Now, those components that need users private data may be kept at OSN site or at some trusted third-party site and flow of information among trusted and untrusted components can be monitored as per the information flow profile of a TPA.}
  \end{enumerate}
 \par Currently, most of the OSNs support a large number of  TPAs that are hosted by various third-party servers and are governed by entities other than OSN providers. Before using any TPA, users need to install it while providing permissions to access the requested data described as follows:
 \subsection{ TPA Installation and Authorization}
\begin{figure}[!ht]
 	\begin{center}
 		\includegraphics[scale=0.3]{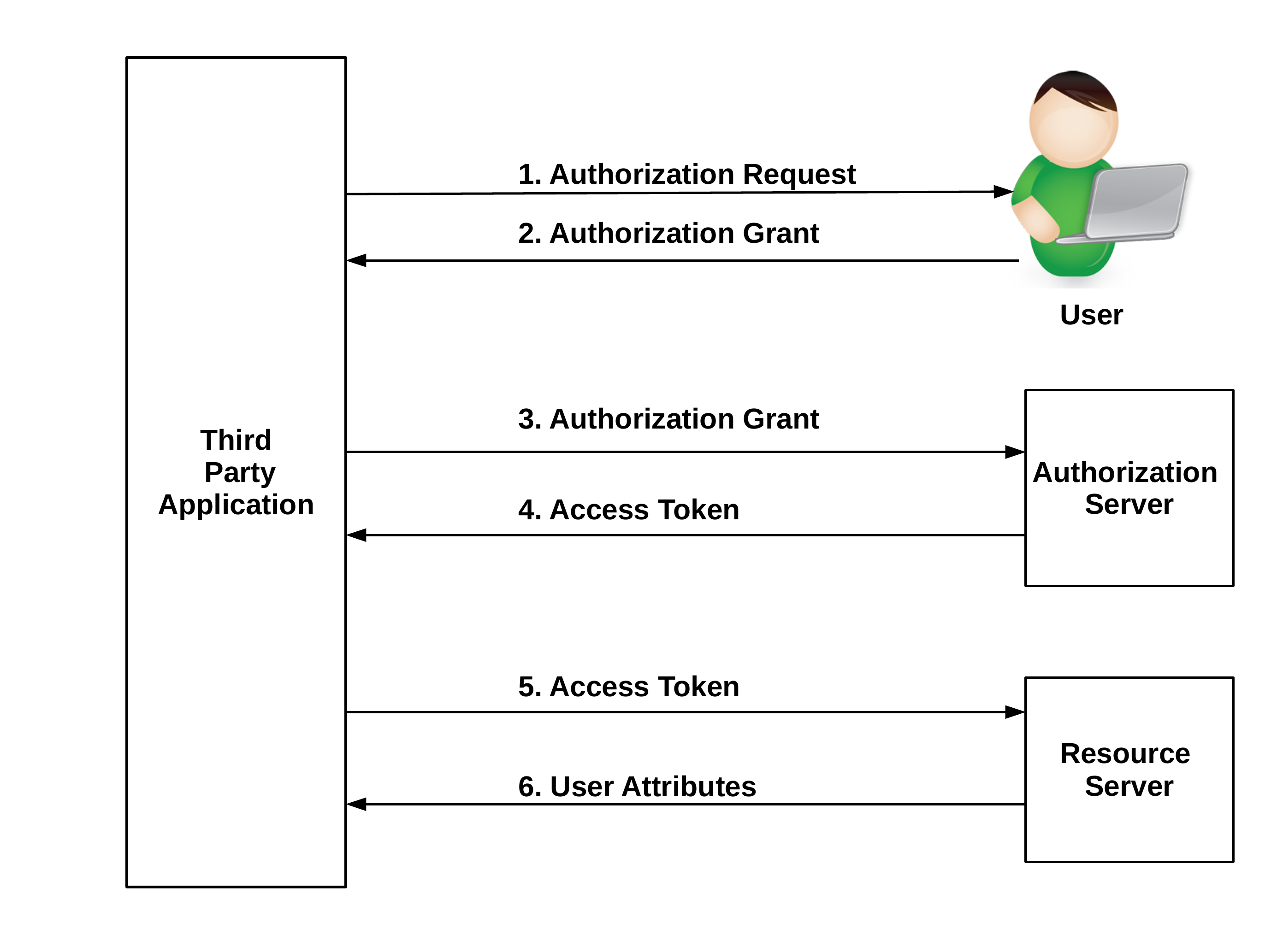}
 		\caption{Abstract Protocol Flow for OAuth 2.0}
 		\label{fig:oauth2}
 	\end{center}
 \end{figure}

 The existing authorization protocol employed by various OSNs uses OAuth 2.0 Protocol \cite{rfc6749}. The abstract flow of this protocol is shown in Figure-\ref{fig:oauth2}.
 \par According to OAuth 2.0 protocol, to access user data, a TPA needs to get an authorization grant for the resources from the corresponding user (flow 1-2). After getting an access grant from user, the TPA obtains an access token from \textit{Authorization Server} (AS), (flow 3-4). TPA can access the subscriber authorized data by submitting this access token to the \textit{Resource Manager} (RM). The major drawback of this approach is that, it offers only a binary approach (authorize access to all or none of the requested data) for delegating the authorization.    
 \subsection{Existing  TPA Integration Framework for OSNs }
\begin{figure}[!ht]
 	\centering
 	\small
 	\includegraphics[scale=0.35]{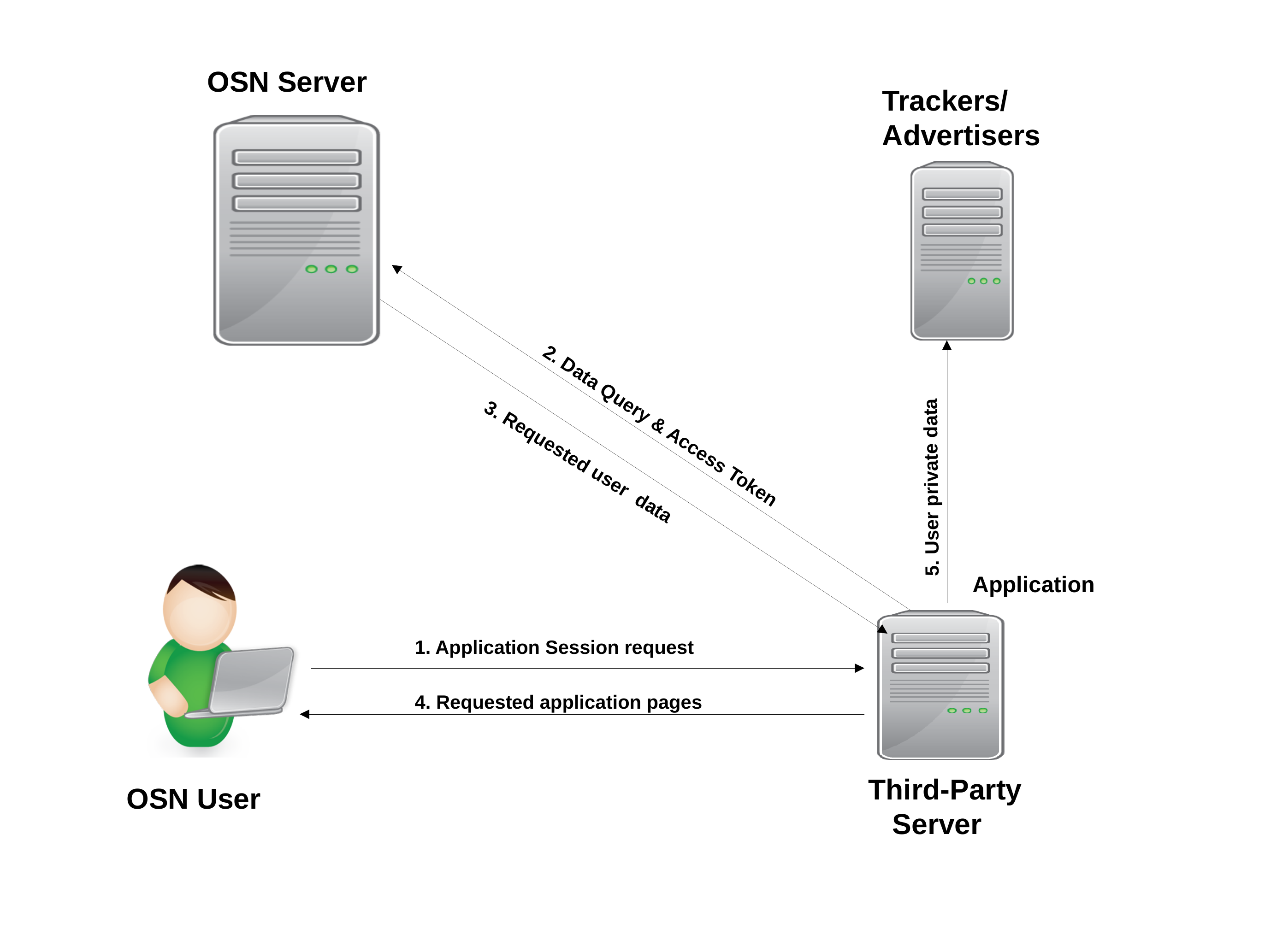}
 	\label{fig:exitingosn-arch}
 	\vspace{-1cm}
 	\caption{Existing TPA Integration Framework}		
 \end{figure}

 Figure-\ref{fig:exitingosn-arch}, shows the existing TPA integration framework for OSNs. Typically, OSN and the corresponding TPAs are hosted at different servers.  
 \par In the beginning of  each session, the corresponding third-party application server sends a data request with the access token to the OSN server. After successful verification of the access token, OSN sends requested data to the TPA server. The TPA server may store this data locally for later use. 
 As a result of this, the subscriber as well as the OSN provider would lose control of the data. In many instances, these TPA servers have been found sharing their subscribers data to various tracking and advertisement agencies resulting in breach of user privacy \cite{Felt:2008}\cite{Chaabane2014}. 
 \par Therefore, it is necessary to control the release of users private data to such applications in order to minimize the risk to users privacy, but without hampering their functionality. Moreover, any solution to the problem described above also need to monitor the functionality of a TPA to avoid any potential information leakage. 
 \section{The Proposed Privacy Preserving Framework} \label{sec:proposedmodel}	
 \subsection{System's Nuts and Bolts}\label{sec:sysmodel}
 OSN can be represented as a directed graph $G =(V,E)$ where $V$, the set of vertices, represents users and third-party applications.
 Let $\mathcal{U} =\{u_1, u_2, ... u_n\}$ be the set of OSN users and $\mathcal{T}$ be the set of all TPAs associated with the OSN. Then, $V=\mathcal{U} \cup \mathcal{T}$. Let $\mathcal{R}= \mathcal{R}_{\mathcal{U}\mathcal{{U}}} \cup \mathcal{R}_{\mathcal{U}\mathcal{{T}}}$ be the finite set of relationships supported by the OSN where $\mathcal{R}_{\mathcal{U}\mathcal{{U}}}$ be the set of relationships between two users,  and $\mathcal{R}_{\mathcal{U}\mathcal{{T}}}$ is the set of relationships between a user and application respectively. $E$ is the set of all directed edges that exist among all the vertices of $G$. Each edge is a 4-tuple $(u, v, r, t)$, where $u, v \in V $, $r \in \mathcal{R}$,
 $t \in [0, 1]$. Here $t$ is a the level of trust that $u$ has on $v$. Further, $t=0$, if $v\in \mathcal{T}$. 
 \begin{table}
 	\centering
 	\small
 	\caption{Important Notations used in InfoRest}
 	\begin{tabular}{||l|l||}
 		\hline \hline 
 		\textbf{ Notation} & \textbf{Meaning} \\
 		\hline \hline
 		$u, ~v$  & OSN users \\
 		\hline
 		$\mathcal{A}$ & TPA \\
 		\hline 
 		$C_{k}^{\mathcal{A}}$ & ${k}^{th}$ component of TPA $\mathcal{A}$  \\
 		\hline
 		$D_u$ &  Set of data item owned by $u$ \\
 		\hline
 		$IFP_{\mathcal{A}}$ & Information flow profile of $\mathcal{A}$\\
 		\hline
 		$C_{I}^{\mathcal{A}}$ & Set of internal components  of $\mathcal{A}$\\
 		\hline 
 		$C_{E}^{\mathcal{A}}$ & Set of external components of $\mathcal{A}$\\
 		\hline 
 		$\mathcal{P}_{i}^{u}$ & The set of policies associated with  $d_{i}^{u}$\\ 
 		\hline \hline
 	\end{tabular}
 	\label{tbl:notations}
 \end{table}
  Now, we define building blocks of InfoRest as follows: 
 \subsubsection{User}\label{subsec:user}
 An OSN user possesses a finite set of attribute-value pairs and objects that define his digital footprint. A subset of these data items may contain sensitive information about the user. The user also has a set of relationships with other OSN users and third-party applications. . 
 \subsubsection{User Data}\label{subsec:userdata}
 A user owns a set of data items (attribute or objects). Let $\mathcal{D}_u =\{d_1^{u},d_2^{u},..,d_k^{u}\}$ be the set of data items belongs to $u \in \mathcal{U}$. We assume that every data item $d_i^{u}$ has a sensitivity level $sl_i^{u}$, depending on the privacy preferences of $u$. The owner $u$ assigns a sensitivity level to  each of the data items possessed by itself. In general, the sensitivity level for the same data item may vary from user to user. We define the following typical \textit{sensitivity levels} for specifying the privacy of subscriber data:   
 \begin{enumerate}
 	\item \textbf{Sensitive Data:} This is the data that contain private information about the user. This personal information may contain \textit{personally identifiable information} that may allow a TPA in establishing the identity of the user. To provide a fine-grained control to user on his data, the sensitive data may be categorized further as follows:  
 	\begin{enumerate}
 		\item \textbf{Highly Sensitive (HS)}: The attributes whose values may be shared with only a specific set of connections such as family friends or similar, with very high trust level can be put in this category. Typically, on a numeric scale attributes with sensitivity level between $0.75$ and $1$ may be put in this category. 
 		\item \textbf{Moderately Sensitive (MS)}: These attribute values may be shared with only a small set of friends that are highly trusted, such as close friends. Typically, on a numeric scale attributes with sensitivity level between $0.5$ and $0.74$ may be put in this category. 
 		\item \textbf{Low Sensitive (LS)}: This data is less sensitive to user and typically can be shared with all the friends. Typically, on a numeric scale attribute with sensitivity level less than $0.5$ may be put in this category. 
 	\end{enumerate}
 	\item \textbf{Non-Sensitive or Public (NS)}: This data do not come into the purview of privacy and do not leak any personal information about the user. This data can be shared with a TPA. The sensitivity value for attributes in this category may be set to $0$ (zero).	
 	
 \end{enumerate} 
 \par A data item $d_i^{u}$ that belongs to a user $u$ is represented by 4-tuple:
 $$d_i^{u} = (id, ~value, ~sl_{i}^{u}, \mathcal{P}_{i}^{u}),$$
 where, \textit{id} refers to the \textit{attribute name} or identifier assigned to  $d_i^{u}$. The $value$ of $d_i^{u}$ refers to the set of values that it possesses. Further, $sl_{i}^{u}$ is the sensitivity level assigned by $u$ to $d_i^{u}$ as per his privacy preferences. And $\mathcal{P}_{i}^{u}$ is the set of policies associated  with it. For example:
\begin{center}
  $ d_i^{u} = (email, \{abc@xmail.com\}, MS,p_i^{u}=\{p_1, p_2 \})$  
\end{center} 

\subsubsection{Third Party Application (TPA)}\label{subsec:tpa}
 We assume that a TPA is a finite set of functional units referred as \textit{components}. A component is a finite set of logically associated \textit{functions}. We divide these components in two categories: \textit{internal components} and \textit{external components}.  Each internal component may access private data as permitted by the corresponding subscriber, but an external component can not access any of the private data. Moreover, internal components are not allowed to share any private data to any external component. 
 \par  InfoRest divides TPAs into two categories: 1) that access only the non-private data of users (App-2 in Figure-\ref{fig:osn-architecturecropped}). This kind of TPAs may be completely hosted outside OSN as they do not access any private data.  2) that access both private and non-private data of users (App-1 in Figure-\ref{fig:osn-architecturecropped}). As described earlier, a typical TPA has two type of components: \textit{internal components} and \textit{external components}.
  \par Formally, a third-party application, $\mathcal{A} \in \mathcal{T}$ is defined by a $3-$tuple i.e.  $\mathcal{A} =(\mathcal{C}_{\mathcal{A}},~\mathcal{D}_\mathcal{A},~ IFP_\mathcal{A})$, where $\mathcal{C}_\mathcal{A}= C_{I}^{\mathcal{A}}\cup C_{E}^{\mathcal{A}}$. $C_I^{\mathcal{A}}$ and $C_{E}^\mathcal{A}$ are the finite set of internal and external components respectively.
  \par $\mathcal{D}_\mathcal{A}$ is the finite set of data items that $\mathcal{A}$ manipulates. These data items may include wall posts, comments, links, a set of memory objects such as variables, buffers, and so on. These data items either store user's private data or intermediate data generated by functions in components. Each object is assigned a \textit{sensitivity label} that reflects its sensitivity with respect to user privacy. The sensitivity of an \textit{intermediate data} item generated by a component is defined as, the highest of the sensitivity values of the set of data items, it is derived from.  
 \par \textit{Information Flow Profile (IFP)} describes normal information uses behavior of $\mathcal{A}$. It contains profile information about all the components of $\mathcal{A}$. It also includes other information such as the application name, domain name, callback URL and a list of user resources whose permission is needed to access the services of $\mathcal{A}$. A typical TPA profile, may be represented as follows:
 \begin{itemize}
 	\item App Title: $\mathcal{A} \in \mathcal{T}$
 	\item Domain name : www.xyz.com
 	\item Callback URL:  www. xyz.com/$\mathcal{A}$/callback.php 
 	\item Required Data List: $\{name, dob, friend\_list \}$
 \end{itemize}
 \begin{table}[!ht]
 	\small
 	\centering
 	\caption{Typical component profile for a TPA components}
 	\vspace{0.3cm}
 	\begin{tabular}{||p{0.6cm}| p{0.69cm}|p{1.5cm}|m{2cm}|m{1.3cm}|>{\centering\arraybackslash}m{2.85cm}||}			\hline \hline 
 		\textbf{ID} & \textbf{Type} &  \textbf{Input } & \textbf{Output / Action} & \textbf{Adjacent Components} & \textbf{External entities} \\
 		\hline \hline 
 		$C_1^{\mathcal{A}}$ & 0  & name, dob  & posts on user's wall & - & www.horoscope.com \\
 		\hline
 		$C_2^{\mathcal{A}}$ & 0  & friend\_list & posts on friends wall & $\{C_{1}^{\mathcal{A}}\}$ & -\\
 		\hline
 		$C_3^{\mathcal{A}}$ & 1  & mouse click & calls click event and $C_1^{\mathcal{A}}$   & $\{click(),$ $C_1^{\mathcal{A}}\}$ & - \\
 		\hline
 		$C_4^{\mathcal{A}}$ & 1  & mouse movement & process mouse movement & - & - \\
 		\hline \hline
 	\end{tabular}
 	\label{{tab:tpaprof}}
 \end{table}

 \par In the above table, for a component with id $C_k^{\mathcal{A}}$, \textit{Input } represents the set of subscriber data, the component needs. The \textit{output } refers to the data, $C_k^{\mathcal{A}}$ generates. Type indicates whether the component is internal $(type=0)$ or external component $(type=1)$. The \textit{adjacent components} is a finite set of components that $C_k^{\mathcal{A}}$ can call.
 The \textit{external entities} refers to the set of external applications or services a TPA component communicates with. 
 \subsection{The InfoRest Design} 
 The following figure shows the design of InfoRest.
 \begin{figure}[!ht]
 	\centering
 	\includegraphics[scale=0.4]{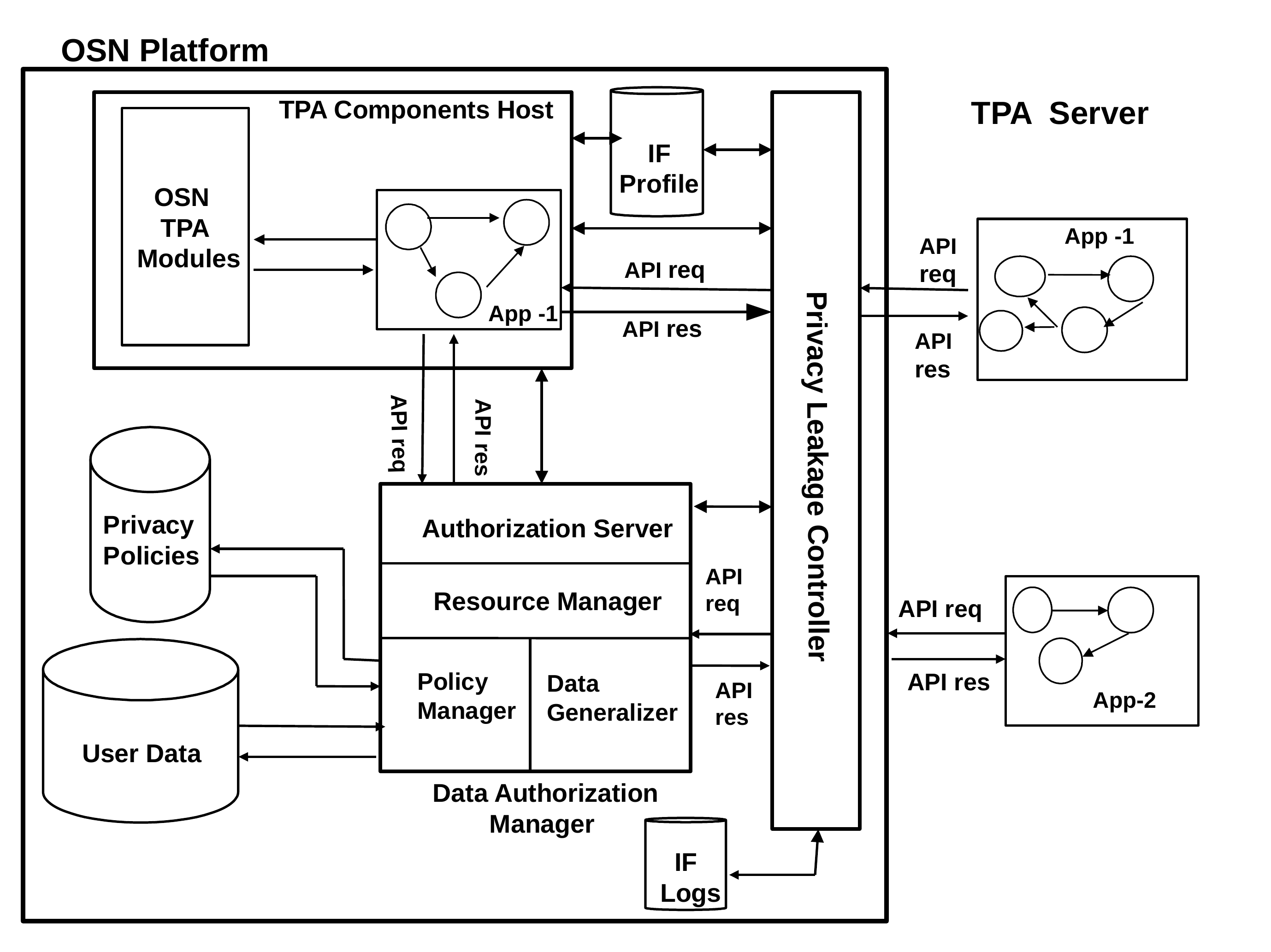}
 	\caption{Proposed architecture of privacy preserving OSN}
 	\label{fig:osn-architecturecropped}
 \end{figure}
 InfoRest comprises of 3 major modules as shown in Figure-\ref{fig:osn-architecturecropped}. These modules are described next. 
 \subsubsection{Data Authorization Manager (DAM)}
 {\em DAM} facilitates users in managing and authorizing their resources. Subscribers may specify their data access policies as per their privacy preferences. Further, InfoRest encodes the policies as per the policy model proposed in Section-\ref{sec:apmodel}. Finally, all authorization policies are stored in a database called \textit{policy database}. DAM enforces these policies during execution of the application. 
 Following are the sub-modules of DAM:
 \par 
 \begin{enumerate}
 	\item \textit{Attribute Generalizer (AG):} 
 	{\em AG} generalizes the value of a user data object while sharing it, if the owner opts for generalization to restrict access of the sensitive data item. The level of generalization may be proportional to the sensitivity label of the data i.e. more the sensitivity, higher would be the generalization to  be used. Any of the available generalization schemes may be used for data generalization \cite{Samarati98,Samarati2001,Li:2008}.
 	\par When a user opts to install a TPA to his  profile, InfoRest offers him a choice to provide generalized values for a subset of attributes required by the TPA.  If the user chooses to use generalization for  some of the attributes, then, he need to provide, generalized values for each of the chosen attribute to DAM. Later, InfoRest shares the generalized values of the resources to serve all the access requests made by the corresponding TPA.
 	\item \textit{Authorization Server (AS)}: It enables a TPA to obtain authorization over a set of user resources. If a user grants access (to a resource), $AS$ issues an access token to the TPA, that enables the TPA to access the resource later. 
 	\item \textit{Resource Manager:} The resource manager keeps records of all the user resources. It accepts all the resource access requests forwarded by the policy manager and responds them accordingly.  
 	\item \textit{Policy Manager:} It enforces the access policies of the subscribers on their respective resources. It receives and evaluates each of the access request made by a TPA. If the access request satisfies the data access policies of the corresponding user, then it forwards the access request to the Resource Manager, and blocks the access request otherwise.  
 \end{enumerate}
  \subsubsection{Privacy Leakage Controller (PLC)}\label{subsec:plc} 
 This module monitors communication among internal and external components of a TPA in order to detect any malicious data flow. It can enable the OSN to ensure that all the data-flows happening among application components are as per the corresponding \textit{information flow assured} by the corresponding TPA developer. Further, it also records all the data flow in \textit{Information Flow (IF) Log} database. So, if any suspicious or abnormal data flow occurs among modules of a TPA, it can inform immediately to OSN provider to block the accessing TPA.
 \subsubsection{TPA Components:}\label{subsec:tpahost}
 Following two types of TPA components are hosted by the OSN.
 \begin{enumerate}
 	\item \textit{OSN TPA Components}:
 	These components may be developed and maintained by OSN provider to provide a common set of services required by TPAs. These components are assumed to be trusted, and therefore may be allowed to access users private data, but they cannot share the accessed data to any TPA module. Such components may include components that offer services such as posting notifications to a user or his friends wall, providing the color palette, calendar, clock and so on.  \\   
 	\item \textit{TPA Internal components }: 
 	An internal component accesses private data of subscribers, so it has to be trusted. Thus, we believe that all internal components of a TPA should be hosted on OSN site, rather than hosting on any third-party server. Furthermore, external components can be hosted at any third-party site, as they do not access any private data of corresponding subscribers. These components only access non-private data (might include sanitized private data) of the subscribers such as mouse movement, key presses etc. They implement application functions that do not depend on any private data. Both internal and external components might communicate with each other by sharing \textit{ non-private data} through functions, as  specified in the \textit{information flow profile} of the application. 
 \end{enumerate}
 
 \subsection{TPA Registration}
\begin{figure}[!ht]	
 	\centering
 	\includegraphics[scale=0.35]{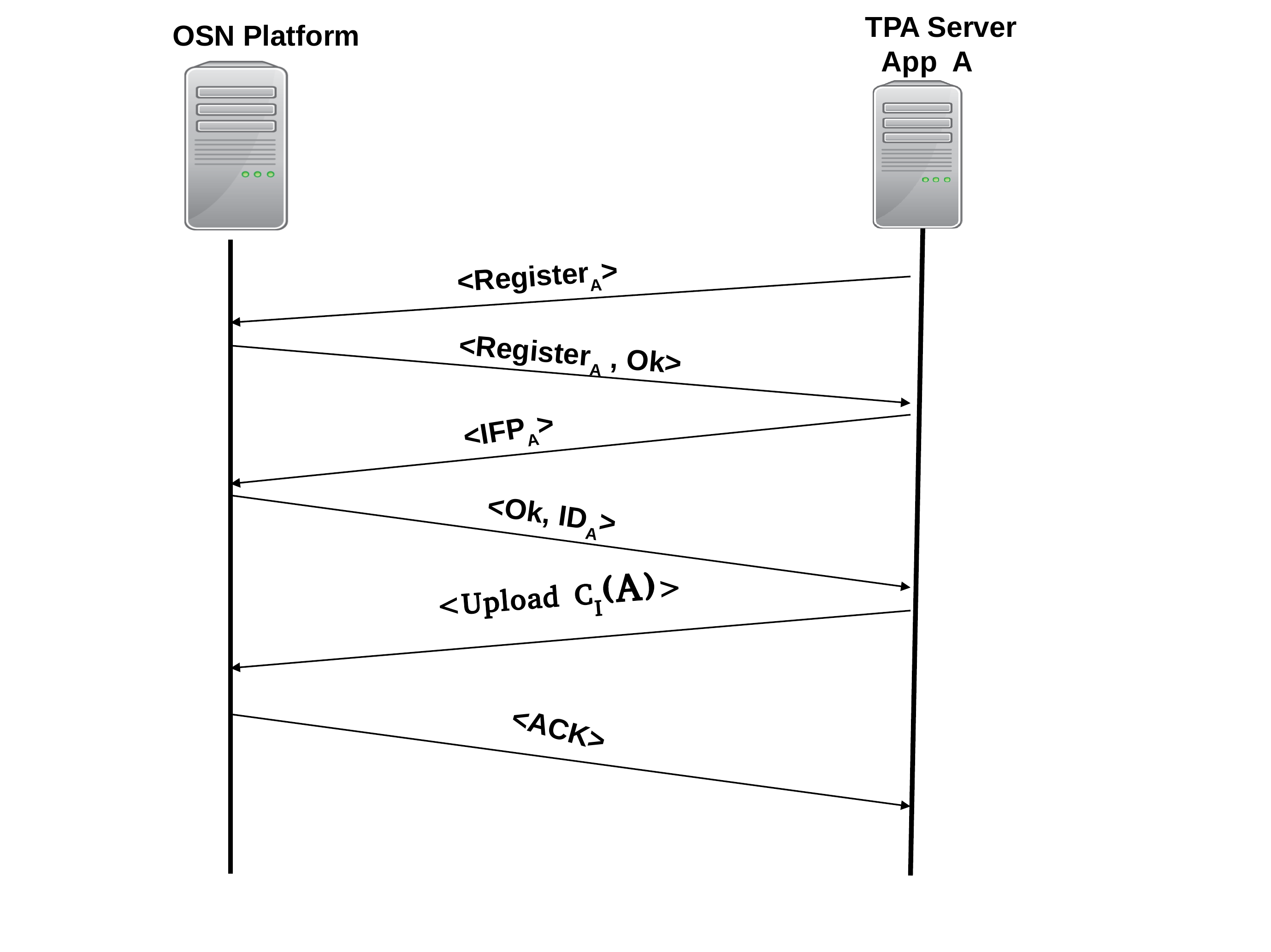}
 	\caption{Application Registration with OSN Service Provider}
 	\label{fig:tparegistration}
 \end{figure}

 Let $\mathcal{A} \in \mathcal{T}$ be an app developed by a third-party developer $\mathcal{D}$. In order to deploy $\mathcal{A}$ at OSN platform, $\mathcal{D}$ needs to register the application with OSN. While registering $\mathcal{A}$, $\mathcal{D}$ shares profile $\mathcal{IFP_{\mathcal{A}}}$ of $\mathcal{A}$ to OSN. $\mathcal{IFP_{\mathcal{A}}}$ includes profile information about each of the components in addition to application name, description, domain and callback URL, and a list of user resources it requires access of. If OSN provider finds that the profile of $\mathcal{A}$ is consistent with its policies, it issues an application-ID to that TPA and stores the profile in the \textit{Information Flow (IF) database}. After that, $\mathcal{D}$ uploads the corresponding internal modules to the OSN platform. Figure-\ref{fig:tparegistration} shows the abstract flow of third-party application registration process with OSN.
 \subsection{TPA Installation and Authorization} As mentioned above, the existing authorization approach is not flexible enough. If a user $u$ wants to use an app, then he has to share the requested resources with the app. Therefore we propose a slight modification in the existing approach. In the beginning,  $u$ is only required to give his willingness to use app $\mathcal{A}$. After that, the app $\mathcal{A}$ forwards this consent to DAM  for getting an access token. DAM performs functions of both the \textit{Authorization Server} and \textit{Resource Manager}. Before issuing any access token, DAM gets the authorization and privacy preferences from the user and issues an access token accordingly. With the help of these access preferences, $u$ can enable DAM and Privacy Leakage Controller to impose restrictions on how a particular attribute may be accessed by $\mathcal{A}$. While providing access preferences $u$ can also request DAM to use data generalization for some resources. In this case, $u$ needs to provide the generalized values for the resources that are to be shared with $\mathcal{A}$. It allows $u$ to have more control over his data.  
 \begin{figure}[!ht]
 	\centering
 	\includegraphics[scale=0.35]{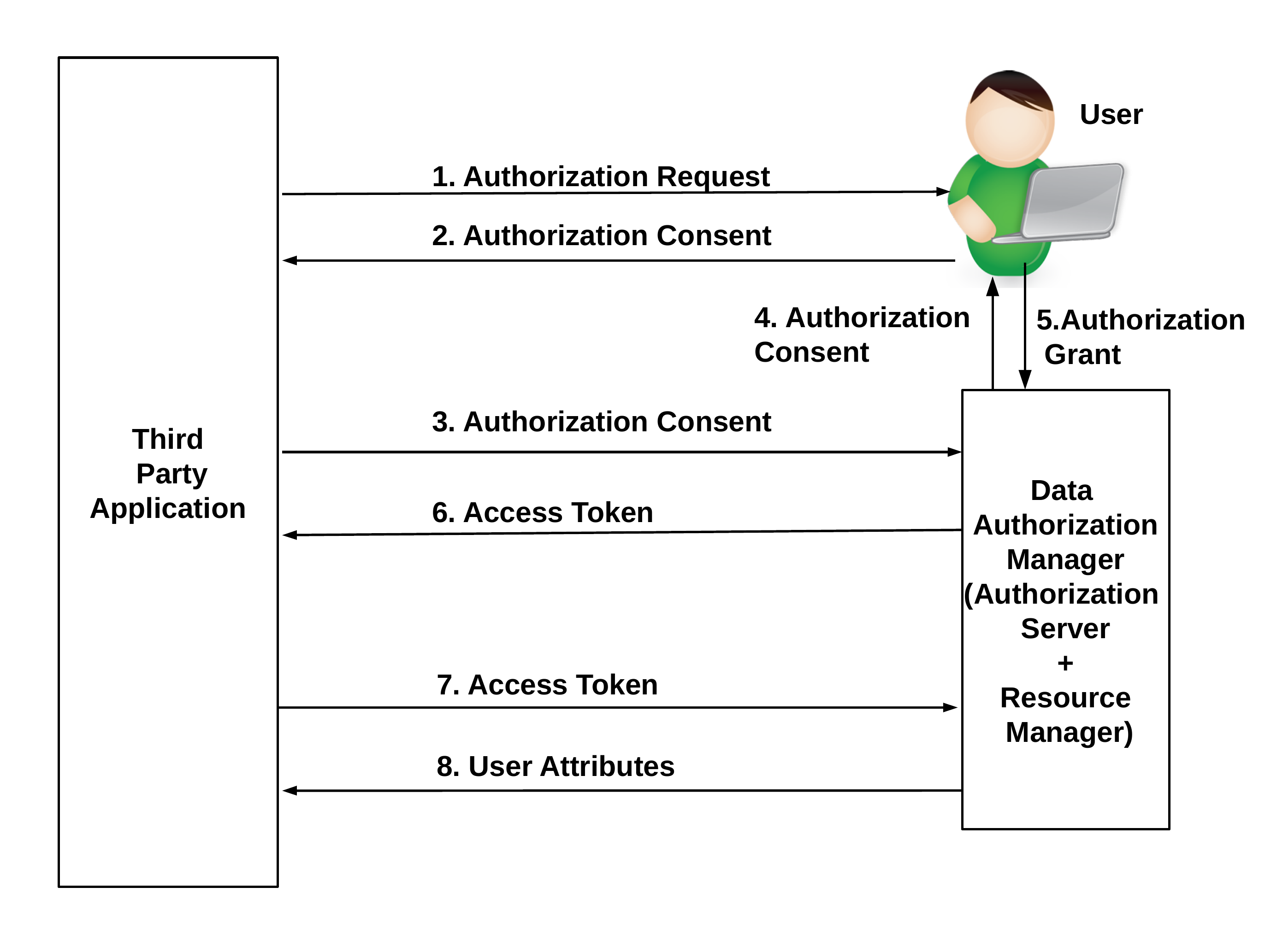}
 	\caption{Authorization by a user to a TPA}
 	\label{fig:tpainstallation}
 \end{figure}
 For the subsequent access, the TPA needs to provide the access token with subsequent access requests. 
 \section{Access Policy Model}\label{sec:apmodel}
 OSN is a \textit{relationship network} where users articulate their communication behavior on the basis of relationships among them. Therefore, Relationship Based Access Control (ReBAC)  would be a suitable choice for controlling data access on OSN platforms. Most of the existing access control mechanisms offered by OSN platforms  are based on user-user relationships. A number of ReBAC mechanisms have also been proposed to control user-user data access \cite{Hu2013}\cite{Cheng2012}\cite{Cheng2014}\cite{Rathore2017}, that are extended to regulate the access of users data in third-party applications \cite{Cheng2013}. We also use ReBAC based access control policy mechanism for TPA in InfoRest.  Our proposed policy language employs \textit{predicate calculus} to specify access conditions for user data as described below. 
 \subsection{Access Policy Specification} 
 \par Let $P_{u}^{\mathcal{A}}$ be the set of policies for TPA $\mathcal{A}$ installed by user $u\in U$. 
 An access control policy $p\in P_{u}^{\mathcal{A}}$  is a 4-tuple as given below:
 $$(target, actions, decision, access~ condition)$$
 where :
 \begin{itemize}
 	\item \textit{target} can be a set of data objects, events or users.
 	\item \textit{actions} is a non-empty finite set of \textit{functions} ($\{f_i\}$) that may be performed on the target.
 	\item  The value of \textit{decision} is either \textit{disallow (=0)} or \textit{allow (=1)}, depending on which the user would be allowed or prevented to perform a function. 
 	\item An \textit{access condition} is a predicate calculus based logical statement that encodes one or more conditions/properties of a target requester  to perform actions specified in the policy. 
 \end{itemize}
 Next, we demonstrate the usability of our policy model in controlling activities of TPAs to access to users data.
 \begin{enumerate}
   	\item  \emph{Policy specification regarding Application Requests and Notifications:} \\
 	Once user $u$ installs TPA $\mathcal{A}$, it may start sending different types of notifications/messages frequently such as advertisements, suggestions about other related apps to $u$ and its friends. Such type of frequent notifications/messages generally irritate users. The proposed model facilitates $u$  to block such type of notifications through the following policy specification.
 	\begin{enumerate}
 		\item To disallow any component of $\mathcal{A}$ to send app suggestions $u$ can use the following policy:\\
 		$[u, \{app\_suggestion\}, 0, [\forall c: is\_componet(c,\mathcal{A}) \wedge installed(u, \mathcal{A})]]$
 		\item To disallow any app suggestions\textbackslash notifications from $\mathcal{A}$, installed by its friends the following policy can be used by $u$:\\
 		$[u, \{app\_suggestion, app\_notification \}, 0, [\forall c, \forall v: is\_component(c, \mathcal{A}) \wedge \neg installed(u,\mathcal{A}) \wedge \\ isfriend(v,u)\wedge installed(v,\mathcal{A})]]$
 		\item  To allow notifications from app $\mathcal{A}$ installed by any of his family member $u$ can use the following policy.\\
 		$[u, \{app\_notification\}, 1, [\forall c,  \exists v: is\_componet(c,\mathcal{A}) \wedge isfamily(v,u)\wedge  installed(v,\mathcal{A})]]$
 		\item To receive updates of TPA  $\mathcal{A}$, installed by $v$, who is a friend of $u$, the following policy may be used:\\
 		$[u,\{app\_updates\}, 1, [\forall c: is\_componet(c,\mathcal{A}) \wedge isfriend(v, u) \wedge installed(v, \mathcal{A})]]$
 	\end{enumerate}
 	\item Policy specification to regulate access to user data:\\
 	User $u$ can regulate access to his private data by TPA $\mathcal{A}$, using  our policy model as follows:\\
 	\begin{enumerate} 
 		\item To allow access of \textit{date of birth} to all internal components of  $\mathcal{A}$ in read mode  without sharing privilege, $u$ may specify following policy:\\
 		$ [-, \{dateofbirth\}, \{read, \neg share \}, 1,  \ [\forall c:  int\_component(c,\mathcal{A}) \wedge installed(u, \mathcal{A})]]
 		$
 		\item To disallow all of the TPAs installed by his friends to read $u$'s friend list and email, $u$ may use the following policy:\\
 		$[-, \{friend\_list, email\}, \{read \}, 0,  [\forall v, \forall c :  isfriend(v,u) \wedge is\_component(c,\mathcal{A}) \wedge installed (v, \\ \mathcal{A})]]$
 		\item  To allow all the external components of app $\mathcal{A}$, to access mouse clicks made by $u$, the following policy may be used:\\
 		$[u, \{mouseclick\}, \{access \}, 1, [\forall c: ext\_component (c,\mathcal{A}) \wedge installed(u,\mathcal{A})]]$
 	\end{enumerate}
     \end{enumerate}
  \subsection{Access Request Evaluation}
 \par An access request is defined by a 3-tuple, (requester, ~target,~actions)
 where, \textit{requester} is an application component that makes the request,  and the other terms (\textit{target} and \textit{actions}) are defined as in access policy specification.
 \par   A TPA component  $C_{k}^{\mathcal{A}}$ sends an access request  $r =(C_{k}^{\mathcal{A}}, d_i^{u}, f_{i}\in F^{\mathcal{A}})$, to get access of a data item $d_i^{u}\in D_u$ to DAM. 
 Consequent upon receiving the request $r$, DAM checks the consistency of $r$ with the profile of $C_{k}^{\mathcal{A}}$.
 
 If the desired action mentioned in $r$ is one of the actions committed in the profile of the component $C_{k}^{\mathcal{A}}$ for the target object, DAM retrieves the corresponding policy $p_{i}^{u}(\mathcal{A})$ and evaluates it. If the request satisfies $p_{i}^{u}(\mathcal{A})$, the access is to be granted and therefore $r$ is forwarded to resource manager to serve the request.
 Otherwise, DAM rejects the request and sends an alert to OSN citing {\em ``suspicious access request"}. 
 \par Further, if $r$ comes from an external component i.e. $C_{k}^{\mathcal{A}} \in C_{E}^{\mathcal{A}}$, then PLC captures the request and verifies its consistency with the profile of $\mathcal{A}$. If $r$ satisfies the consistency requirement specified in the profile, PLC forwards the same to DAM for further evaluation. Subsequently, PM serves $r$ as per the user policy associated with the target object, otherwise denies the request and sends an alert to OSN regarding the unauthorized access attempt made by the component of $\mathcal{A}$. \\
 \emph{Conflict Resolution:} Policies of two or more users may conflict with each other due to different privacy preferences. For example, user $u$ may allow $\mathcal{A}$ to post notification to all his friends wall but user $v$ may not allow $\mathcal{A}$ to send any notification to his friends wall. In this case, the policy conflict would arise for the users that are mutual friends of both $u$ and $v$. To resolve any policy conflict, we adopt the following conflict resolution mechanism:
 \begin{enumerate}
 	\item If a policy conflict happens, the policy specified by the \textit{target user} for the data item gets \textit{precedence}.
 	\item If the target user does not specify any policy for the data item, then the policy of the most trusted contact of the target gets precedence. 
 \end{enumerate}
 \subsection{Model Verification}
 We employ \textit{Answer Set Programming (ASP)}\cite{Lifschitz:2008:ASP} to analyze the correctness of authorization of our policy model. 
 We created an ASP program  $\Pi_{KDB} \cup \Pi_{Policy} \cup  \Pi_{query}$, where $\Pi_{KDB}$ is a knowledge database. $\Pi_{KDB}$ represents a small OSN with users, TPAs, connection among users and TPAs, and a set of data items possessed by users.  All of this information is represented using facts in ASP. A fact is an ASP rule with empty body. $\Pi_{Policy}$ contains user policies encoded as ASP rules. Further, $\Pi_{query}$ encodes the the  access requests. We represent a user policy using an ASP rule whose body encodes the access conditions and head represents the decision. A sample of KDB would be is as follows.
 \begin{itemize}
 	\item user (tom; jerry; adam; ajay; meena; akash; jitendra).
 	\item tpa (mario; minesweeper; spade).
 	\item has\_installed (meena, mario; adam, minesweeper).
 	\item data\_item (name; email; gender; friendlist; photos). 
 	\item  action (post; read; write; transfer; notification).
 	\item data\_value (ajay, email, ajayatiitpacin; ajay, gender, male).
 	\item  family\_friend (ajay, jitendra). 
 	\item colleagues (adam, meena).
 \end{itemize}
 User policies are represented in  the form of ASP rules. Some of the rules are given below:
 \begin{enumerate}
 	\item Rule $R_1$ enables a user $U$ to allow, only the TPAs installed by him, to perform the permitted actions only on a data item D. \\
 	$R_1$: giveAccess2App(D, U, X, T) :- not - allowedtoTPA(U, X, D, T), request (U, X, D, T), user (U), action (X), data\_value (U, D, V), data\_item (D), has\_installed (U, T), tpa(T).
 	\item Rule $R_2$ permits a TPA that is installed by a colleague of $U$, to post notifications on his \textit{wall}. A specific name of colleague and TPA also may be provided by $U$ as well if needed. \\
 	$R_2$: giveAccess2App (wall, U, notification, T) :- not - allowedtoTPA (U, notification, wall, T), request (U, notification, wall, T), colleagues (U, U1), user (U), user (U1), action (notification), data\_value (OW, wall, V), has\_installed (U1, T), data\_item (wall), tpa (T).
 	\item Rule $R_3$: disallows all the apps installed by a user's friends to send any app suggestion to the wall of a user.\\
 	$R_3$:disallowAccess2Comp (wall, U, appsuggestion, C, T) :- not allowedtoComp (U, appsuggestion, wall, C, T), request (U, appsuggestion, wall, C, T), user (U), action(appsuggestion), data\_item (wall), tpacomponent(C), iscomponentof (C, T), tpa (T), has\_installed (U1,T), friend(U1,U), user(U1).
 	\item Rule $R_4$ enables a user to disallow all TPAs to access his \textit{dateofbirth}.\\
 	$R_4$: disallowAccess2Comp (dateofbirth, U, access, C, T) :- not allowedtoComp (U, access, dateofbirth, C, T), request (U, access, dateofbirth, C, T), user (U), action (access), data\_item (dateofbirth), tpacomponent (C), iscomponentof (C, T), tpa (T).
 	\item Rule $R_5$ enables a user to allow an external component of a TPA  \textit{linktrust} to read his \textit{mouse click}.\\
 	$R_5$: allowAccess2Comp (mouseclick, U, read, C, linktrust) :- not - allowedtoComp (U, read, mouseclick, C, linktrust), request (U, read, mouseclick, C, linktrust), user (U), action(read), data\_item (mouseclick), tpacomponent(C), iscomponentof (C, linktrust), isexternalComp(C, linktrust), tpa (linktrust).
 	
 \end{enumerate} 
 \par We used Gringo-4.5.4 \cite{gringo} as \textit{grounder} and Clasp-3.1.5 \cite{gekasc12c} as \textit{answer set solver} on a machine having i7-(2.70 GHz) Intel processor, 4GB RAM, and 64-bit Ubuntu 16.10 platform. To verify if a given access request is satisfied, we submit a query to our ASP logical model encoding the request as an ASP fact. If the required resource policy is satisfied, then the returned answer set contains at least one ASP fact indicating the access grant. If the request does not satisfy the corresponding user policy then the fact negating the grant appears in the answer set.  We evaluated various kind of policies and established correctness of authorization encoded by those policies. 
 \par We verified correctness of the authorization encoded with our policy model by querying $\Pi_{KDB} \cup \Pi_{Policy}$.  In order to establish the correctness of the model, we checked for two things, the \textit{oversharing} and \textit{undersharing}. \\
 \emph{Checking Oversharing:} Oversharing means that one or more unauthorized TPA components get access to a user resource. To check oversharing, we submit a query on behalf of each of the components, not authorized to access a resource. If the resulted answer set contains facts, indicating that any of the unauthorized user is permitted to access the resource then oversharing is present. During analysis, we checked for oversharing for most of the policies we specified, and did not find any instance of oversharing. \\
 \emph{Checking Undersharing:} Undersharing means that one or more authorized components do not get access to a resource. To check undersharing during the analysis, we issued query to $\Pi_{KDB} \cup \Pi_{Policy}$ on behalf of authorized components for most of the permitted policies in $\Pi_{KDB}\cup \Pi_{Policy}$.  We retrieved and observed each of the answer sets corresponding to these queries. During analysis, we found that all of these answer sets contained facts (derived during grounding and solving process) for each of the authorized user. It means that no authorized component was denied access to a resource by our policy model. 
 \section{Discussion}\label{sec:discussion}
 InfoRest offers flexible attribute sharing with TPA. Moreover, instead of using \textit{all-or-nothing approach}, our framework offers choice of sharing  generalized values for their sensitive attributes. The Table-\ref{tbl:comparisonprev} provides a comparison of InfoRest with some of the popular solutions among the existing solutions.
 
 \begin{table}
 	\centering
 	\small
 	\caption{Comparison with existing solutions}
 	\vspace{.05cm}
 	\begin{tabular}{|l|c|c|c|c|c|c|c|}
 		\hline \hline
 		-  &\cite{Viswanath2012} &\cite{Anthonysamy:2012} &\cite{Shehab2012}&  \cite{Egele2012} & \cite{Cheng2013}  & \cite{Tomy2016} & InfoRest \\
 		\hline \hline
Transparency & \checkmark & -  & \checkmark & - & \checkmark & \checkmark & \checkmark\\
 		\hline
 		Information Flow Control & \checkmark & - & - & - & - & - & \checkmark \\
 		\hline
 		Data Generalization & - & - & \checkmark & - &- & - & \checkmark \\
 		\hline
 		User Specified Privacy Preferences & - & \checkmark & \checkmark & \checkmark & \checkmark & - &\checkmark \\
 		\hline
 		User-TPA Policy Model & - & - & - & - &  \checkmark & - & \checkmark \\ 
 		\hline 
 		Components based TPA Model & \checkmark &- & -& - & \checkmark & - &\checkmark \\
 		\hline
 		Zero Store & - & - & - & - & - & -  & \checkmark \\
 		\hline	
 		Policy Conflict Resolution & - & - & - & - & - & - & \checkmark \\
 		\hline	
 	\end{tabular}\label{tbl:comparisonprev}
 \end{table}  
{Since relation is the basis of all  social networks, a ReBAC model is more suitable. Therefore, we choose to develop a ReBAC model for policy specification in InfoRest. It employs \textit{predicate calculus} to express access conditions}.
Only Cheng's model \cite{Cheng2013} among the existing models, has a \textit{User-TPA policy model} but lacks a conflict resolution mechanism. It uses ReBAC policy model for access control for third-party applications,  but it does not allow a user to express policies containing phrases \textit{for all} or \textit{for some}. Also, it enables users to specify policies which could contain ``there exist" or similar  phrases. Therefore, the policies like those discussed below, can not be expressed in the model \cite{Cheng2013}, but that can be expressed in InfoRest: \\
 \textit{Example1:} Suppose $u$ wants to receive  app notifications from all the apps installed by at least one of his family members. In our model the following policy may be used:\\
 $(u, \{app\_notification\}, 1, [\forall c~ \forall \mathcal{A}~ \exists v: is\_componet(c,\mathcal{A}) \wedge isfamily(v, u) \wedge installed(v, \mathcal{A})])$\\
 \textit{Example2:} Suppose $u$ wants to receive  app notifications from all the apps that has been installed by all of his family members. In our model the following policy allow this:\\
 $(u, \{app\_notification\}, 1, [\forall c~ \forall \mathcal{A}~ \forall v: is\_componet(c,\mathcal{A}) \wedge isfamily(v, u) \wedge installed(v, \mathcal{A})])$ \\
 \textit{Example3:} Suppose $u$ wants to allow  score updates from all the apps that has been installed by any of his classmates but not by any of his co-worker. In our model the following policy allow this:\\
 $(u, \{score\_update\}, 1, [\forall c~ \forall \mathcal{A}~ \exists v ~\nexists y: is\_componet(c,\mathcal{A}) \wedge isclassmate(v, u) \wedge installed(v, \mathcal{A}) \wedge  iscoworker(y,\\ u) \wedge \neg installed(y, \mathcal{A})])$ 
 
 \par The main advantage of InfoRest over the existing models is that it does not allow private data of subscribers to move outside of OSN platform. This enables users to have full control over their private data, and denies any control to TPA provider on the data. InfoRest also enable OSN provider to ensure \textit{principle of least privileges} against TPAs by monitoring the flow of private data among TPA components.  Another advantage of our model is that it allows users to share a generalize value for any of the private data as per their sensitivity level. This allows users to reduce the risk of disclosure of their private attributes through TPAs. 
\section{Conclusion}\label{sec:conclusion}
 In this paper, we proposed a new framework called InfoRest, for OSNs to restrict leakage of the subscriber's private information to the external TPAs. Our model uses ReBAC policy model to control access of the user data by application components.  We have verified the correctness of the authorization encoded by our policy model with the help of an ASP logical model. 
 \par One of the major advantage of our framework is that, it does not allow any of the private data of a user to move outside of the  OSN platform. Our model also provides a ReBAC policy model to control access of sensitive data by TPAs. Furthermore, the proposed framework allows a user to choose the option for sharing a generalized value of the private data to avail the service provided by TPA. As future work, we would like to develop a flexible and automatic attribute generalization method and a prototype of our scheme. 


\bibliographystyle{unsrt}  


\bibliography{ref}

\end{document}